\documentclass[9pt,twocolumn,twoside]{opticajnl}
\journal{opticajournal} 

\usepackage{graphicx}
\usepackage{dcolumn}

\usepackage{graphicx}
\usepackage{soul}
\usepackage{dcolumn}
\usepackage{bm}

\usepackage[utf8]{inputenc}
\usepackage[T1]{fontenc}
\usepackage{mathptmx}
\usepackage{etoolbox}
\usepackage{lineno,comment,siunitx,ulem,xcolor,textgreek}
\usepackage{subcaption}

\usepackage{bm}
\setboolean{shortarticle}{true}

\title{Few-cycle THz Pulse Generation in DSTMS Crystal Pumped by a 8.3-MHz Amplified Mamyshev Oscillator}

\author[1]{Francesco Canella}
\author[2,3*]{Dario Giannotti}
\author[4]{Sara Pizzurro}
\author[4]{Riccardo Gotti}
\author[5,6]{Lorenzo Mosesso}
\author[5,6]{Stefano Lupi}
\author[4]{Antonio Agnesi}
\author[4]{Federico Pirzio}
\author[1,3]{Gianluca Galzerano}

\affil[1]{Istituto di Fotonica e Nanotecnologie, Consiglio Nazionale delle Ricerche, Piazza Leonardo da Vinci 32, 20133 Milano, Italy}
\affil[2]{Dipartimento di Fisica, Politecnico di Milano, Piazza Leonardo da Vinci 32, 20133 Milano, Italy}
\affil[3]{Istituto Nazionale di Fisica Nucleare, Sezione di Milano, Via Celoria 16, 20133 Milano, Italy}
\affil[4]{Dipartimento di Ingegneria Industriale e dell’Informazione, Università di Pavia, Via Ferrata 5, 27100 Pavia, Italy}
\affil[5]{Dipartimento di Fisica, Università degli Studi di Roma ”La Sapienza”, Piazzale Aldo Moro 5,00185, Rome, Italy}
\affil[6]{Istituto Nazionale di Fisica Nucleare, Sezione di Roma, P.Le Aldo Moro 2, 00185 Rome, Italy}

\affil[*]{dario.giannotti@polimi.it}

\begin{abstract}
Mamyshev oscillators are an emerging class of ultrafast fiber lasers that support exceptionally broadband spectra and few-femtosecond pulse durations, making them well-suited for nonlinear frequency conversion.
Despite this potential, THz generation using Mamyshev oscillators has not been demonstrated to date.
In this work, we report the generation of THz few-cycle at 8.3~MHz repetition rate via optical rectification of a 31-fs pulse duration, 1-W average power amplified Mamyshev oscillator in a 190-µm-thick DSTMS organic crystal.
We measured a THz average power of 40 µW and a spectral bandwidth of 4~THz. To further investigate the advantage of combining Mamyshev oscillator and organic crystals for THz generation at multi-MHz repetition rate, we compared the THz pulses with those generated using a conventional inorganic 500 µm-thick GaP crystal, obtaining comparable bandwidth, but 20 times lower power with respect to DSTMS.

\end{abstract}

\setboolean{displaycopyright}{false} 

\begin{document}

\maketitle

\begin{figure}[ht]
\centering
\includegraphics[width=\linewidth]{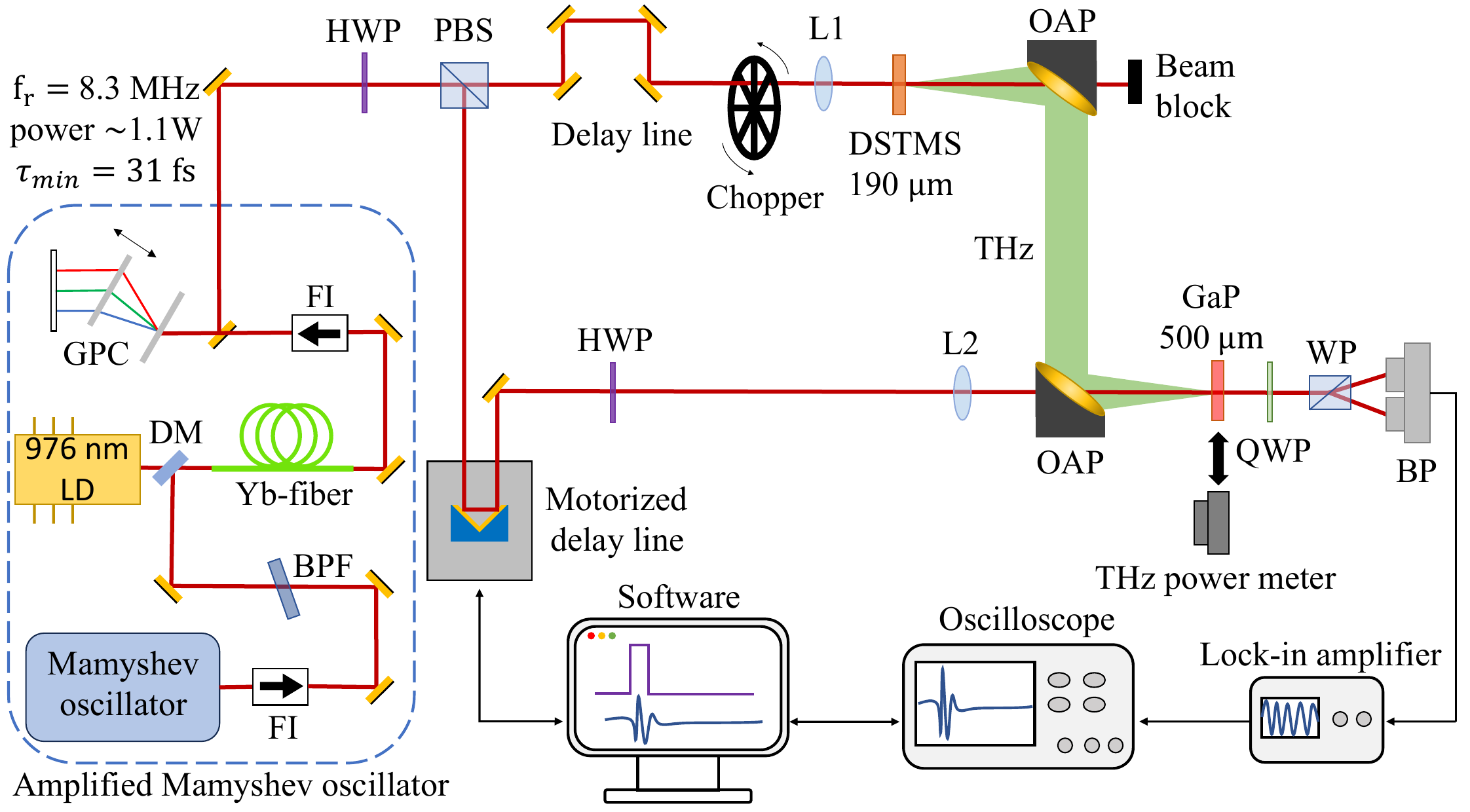}
\caption{Experimental setup schematics. BPF: band-pass filter; FI: Faraday isolator; DM: dichroic mirror; GPC: grating pair compressor; H/QWP: half/quarter-wave plate; PBS: polarizing beam splitter; L1 and L2: lenses $f = \SI{200}{\milli\meter}$; OAP: off-axis parabolic mirror; WP: Wollaston prism; BP: balanced photodiode.}
\label{fig:setup}
\end{figure}

\begin{figure}[ht]
\centering
\includegraphics[width=\linewidth]{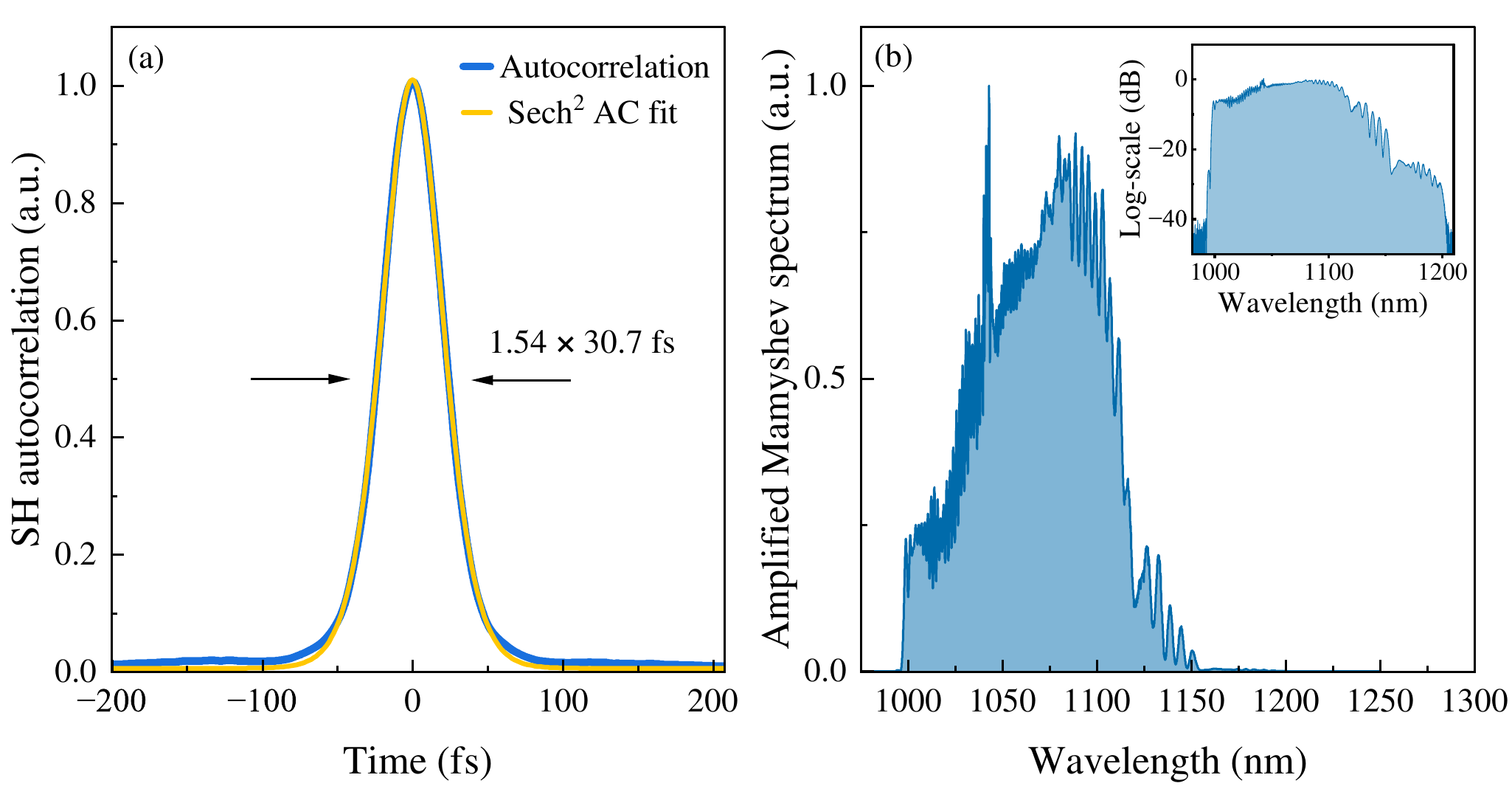}
\caption{Amplified Mamyshev oscillator: (a) SH intensity autocorrelation with $\text{Sech}^2$ fit, and (b)  spectrum in linear and log scale (inset).}
\label{fig:Mamyshev}
\end{figure}

Ultrafast lasers are today the cornerstone of several scientific fields and applications, such as ultrafast spectroscopy, laser micromachining, and nonlinear optics \cite{Reid2016}.
In recent years, ultrafast ytterbium-doped fiber lasers have become increasingly popular due to their stability, compactness, and robustness compared to solid-state-lasers.
Mamyshev oscillators (MOs) in fiber represent a cutting-edge advancement in contemporary ultrafast fiber laser technology at 1~μm wavelength, offering excellent performance in pulse energy (>100~nJ), very large nonlinear phase (on the order of 100$\pi$), and high-quality compressibility down to $\sim$30~fs duration \cite{Sidorenko2018, Liu2019}.
Furthermore, MOs are less sensitive to environmental noise with respect to other common architectures based on nonlinear polarization evolution (NPE), on the use of SESAMs or of nonlinear amplifier loop mirrors (NALMs) \cite{Chong2015}.
One of the main challenges in MOs is initiating the mode-locking regime, which is often achieved through complex techniques such as injection of a short picosecond pulse produced by another mode-locked oscillator or pump modulation.
Only recently, a sub-ns passively Q-switched (PQS) external starter has been successfully employed as a simple, robust, and cost-effective solution to reliably initiate the mode-locking regime, even in high-gain configurations based on large-mode-area (LMA) fibers \cite{Gotti2024a, Gotti2024b}.
The recent discovery of the pulse propagation regime known as gain-managed nonlinear amplification (GMNA) has enabled further significant advances \cite{Sidorenko2019}.
Indeed, a proper pulse amplification is essential to transfer the attractive features of MOs to applications where high energy is required, overcoming the intrinsic limits of traditional approaches like chirped-pulse-amplification (CPA).
The peculiarity of GMNA regime is the interplay of positive dispersion and spectral gain evolution with pulse amplification along the fiber to reach a spectral bandwidth beyond the gain bandwidth of the amplifier itself.
We recently showed that by selecting with a band-pass filter (BPF) the proper seeding wavelength within the spectrum of the MO, it is possible to optimize the performance of the GMNA amplifier in terms of suppression of the compressed pulse pedestal and minimization of the pulse duration, resulting in a higher pulse peak power \cite{Pizzurro2025} with respect to \cite{Gotti2024a, Gotti2024b}.
These features make GMNA amplifiers seeded by MOs particularly well-suited for nonlinear optics experiments, such as mid-infrared optical parametric generation and laser-driven terahertz waves (THz) generation via optical rectification (OR) in $\chi^{\text(2)}$ crystals \cite{Sukeert2023, Fulop2020}.
Nevertheless, to the best of our knowledge, no OR experiments with MOs have been reported to date.
This application of MOs is of great interest, since ultrafast laser-driven few-cycle THz pulse sources are now widely recognized as essential instruments in many areas of fundamental and applied research, as well as technological and industrial applications \cite{Jepsen2011}.
Photoconductive antennas offer an alternative table-top approach, but a significant gap remains between prototypes and commercial devices, particularly at MHz repetition rates, where carrier-induced heating in materials limits average output power and long-term stability\cite{Khalili2025}.
In contrast, $\chi^{\text(2)}$ nonlinear crystals are widely available on the market. A wise choice of both the crystal and the pump laser (wavelength, pulse duration, pulse energy, and repetition rate) is critical for defining the achievable THz bandwidth, pulse energy, and data-acquisition speed.
All these parameters need to be balanced to maximize the dynamic range (DR) and the signal-to-noise ratio (SNR) of THz spectroscopic measurements \cite{Jepsen2005}.
High bandwidth (> 4~THz) is usually obtained at the expenses of the DR with thin crystals, while high energy is achieved at the expenses of repetition rate of the pump laser, which can degrade SNR unless acquisition times are extended significantly \cite{Mansourzadeh2023}.
A great effort has been made to solve the latter problem by making high-average power THz sources with repetition rates exceeding the MHz.
Average power levels on the order of mW at repetition rate exceeding tens of MHz have been demonstrated with OR of thin LiNbO$_3$ (lithium niobate, LN) plates o gallium phosphide (GaP) inside active and passive optical cavities \cite{Hamrouni2021, Wang2023, Suerra2025}.
Cavities boost the pulse energy at high repetition rates, but they require specific know-how and electro-optical components to work properly.
For this reason, for future technology scalability, single-pass OR still remains the simplest and preferable option.
In this context, organic crystals have emerged as excellent alternative candidates to traditionally used inorganic crystals (mainly GaP, LN, ZnTe, and GaSe), since they combine the advantages of a broad generation bandwidth in a simple collinear geometry with high conversion efficiency.
The most common organic crystals used for OR pumped by near-infrared lasers are DSTMS (4-N,N-dimethylamino4’-N’-methyl-stilbazolium 2,4,6 trimethylbenzenesulfonate), BNA (N-benzyl-2-methyl-4-nitroaniline), OH1 (2-[3-(4~hydroxystyryl)-5,5dimethylcyclohex-2-enylidene] malononitrile), HMQ-TMS (2-(4-hydroxy-3-methoxystyryl)-1-methilquinolinium 2,4,6-trimethylbenzenesulfonate), and DAST (4-N,N-dimethylamino-4’-N’-methyl-stilbazolium tosylate).
Comprehensive information on organic crystals can be found in Ref. \cite{Jazbinsek2019}.
Note that the high conversion efficiency is mainly due to a large effective nonlinear coefficient $d_{\text{eff}}$, which contributes quadratically to the conversion efficiency \cite{Fulop2020}.
e.g., GaP have a $d_{\text{eff}} = \SI{24}{\pico\meter\per\volt}$, while DSTMS have $d_{\text{eff}} = \SI{230}{\pico\meter\per\volt}$ \cite{Mansourzadeh2023}.
Among inorganic crystals, LN is a special case, since is has an high $d_{\text{eff}} = \SI{160}{\pico\meter\per\volt}$, but it is out of phase matching on almost all the near infrared region.
As a drawback, organic crystals are sensitive to high average power: specific substrates or chopping are often required to avoid thermal damage
\cite{Mansourzadeh2021}.
The literature reports several notable examples of high-repetition-rate THz generation using organic crystals pumped by Yb lasers.
Among them, in Ref \cite{Buchmann2020} Buchmann et al. demonstrated the generation of mW-level quasi-single-cycle THz pulses at 10-MHz repetition rate via OR in HMQ-TMS, pumped by an Yb-fiber laser whose pulses were recompressed by a multipass cell to 22~fs.
In 2021, Mansourzadeh et al. proved 0.95~mW of THz at 13-MHz repetition rate using a high-power thin-disk laser and a BNA.
With a 100-kHz repetition rate, K. Wang et al. \cite{KangWang2024} recently demonstrated strong field THz using a DSTMS crystal.
However, a study of OR in DSTMS with Yb lasers at MHz repetition rate is still unpublished.
In this work, we report THz generation via OR in DSTMS and GaP crystals exploiting an amplified Yb-doped fiber Mamyshev oscillator.
Our laser has a repetition rate of 8.3~MHz, it is externally triggered through PQS, and operates in the GMNA regime.
Our source delivers high-quality pulses compressible to 31 fs without side lobes, reaching peak powers up to ~4 MW and allowing average THz powers in the tens of \SI{}{\micro\watt} range.

The setup of THz generation is shown in Fig.\ref{fig:setup}. The BPF selects a $\sim$3-nm FWHM portion of the $\sim$25-nm wide MO spectrum to seed the GMNA amplifier with the resulting $\sim$ 2-mW average power pulse train.
The laser is amplified to \SI{1.1}{\watt} in a Yb-doped fiber amplifier pumped by a 7-W multimode diode at \SI{976}{\nano\meter}.
After amplification, the pulses are compressed to a minimum duration of \SI{31}{\femto\second} by a grating pair (1000 grooves/mm, single pass transmission >97\% at 1040~nm) in the Treacy configuration.
The second-harmonic (SH) intensity autocorrelation and the spectrum of the amplified pulses are shown in Fig.\ref{fig:Mamyshev}, in panels (a) and (b), respectively.
The autocorrelation is well fitted by a $\text{Sech}^2$ profile, which highlights the excellent quality of the pulse.
The optical spectrum has a FWHM $>\SI{80}{\nano\meter}$, and in log-scale it extends from $\sim$ 1~µm to above 1.2~µm  (SNR $\approx 40$~dB).
An extensive description of the GMNA seeded by a MO can be found in Ref. \cite{Pizzurro2025}.
To generate THz, the amplified pulses are selected in polarization with a half-wave plate (HWP) and a polarizing beam splitter (PBS).
Then, the beam passes through a mechanical chopper with selectable frequency and duty cycle and is focused by a 200-mm focal length lens into a DSTMS organic crystal for single-pass collinear OR. For comparison, the DSTMS can be substituted by a GaP placed in the same configuration.
The DSTMS (SwissTHz) is uncoated and it is 190-µm thick.
It is placed on a rotating mount to optimize the angle of phase matching.
The organic crystal is deposed on a millimetric silica substrate.
The crystal thickness is chosen to stay within the coherence length up to 5~THz.
Inside the crystal, the dimension of the waist of the laser is $w_{\rm IR}=\SI{100}{\micro\m}$.
The THz radiation diverges from a dimension of $w_{\rm THz} = w_{\rm IR} /\sqrt{2} \approx \SI{70}{\micro\m}$ and is collected by a protected gold off-axis-parabolic mirror (OAP) of focal length and diameter of 50.8~mm. The infrared beam is discarded by a 3-mm hole in the center of the OAP (without significant THz losses).
We estimate that the divergence of the THz beam (especially below 1~THz) causes a loss of $\sim$50-55\% of power \cite{Faure2004}.
The first OAP collimates the THz, while a second identical OAP focuses the beam for applications and characterization. 
The distance between the two OAPs is twice their focal length (101.6~mm).
At the focal point, we performed the characterization of the generated pulses in terms of power (using a THz power meter) and in terms of pulse shape and spectrum through electro-optical sampling (EOS) \cite{Benea-Chelmus2025}.
For power characterization, we used a pyroelectric power meter (Ophir RM9-THz).
The background floor is 200~nW.
The sensor imposes a chopping frequency of the beam at 18~Hz, with duty cycle 50\%. To prevent residual scattered infrared light from reaching the THz detector, we shielded it with two calibrated sheets that completely suppress the infrared while transmitting 26\% of the THz.
The waist of THz beam at the second OAP focus, measured using the knife-edge method, is $\sim$420~μm.
\begin{figure}[t!]
\centering
\includegraphics[width=\linewidth]{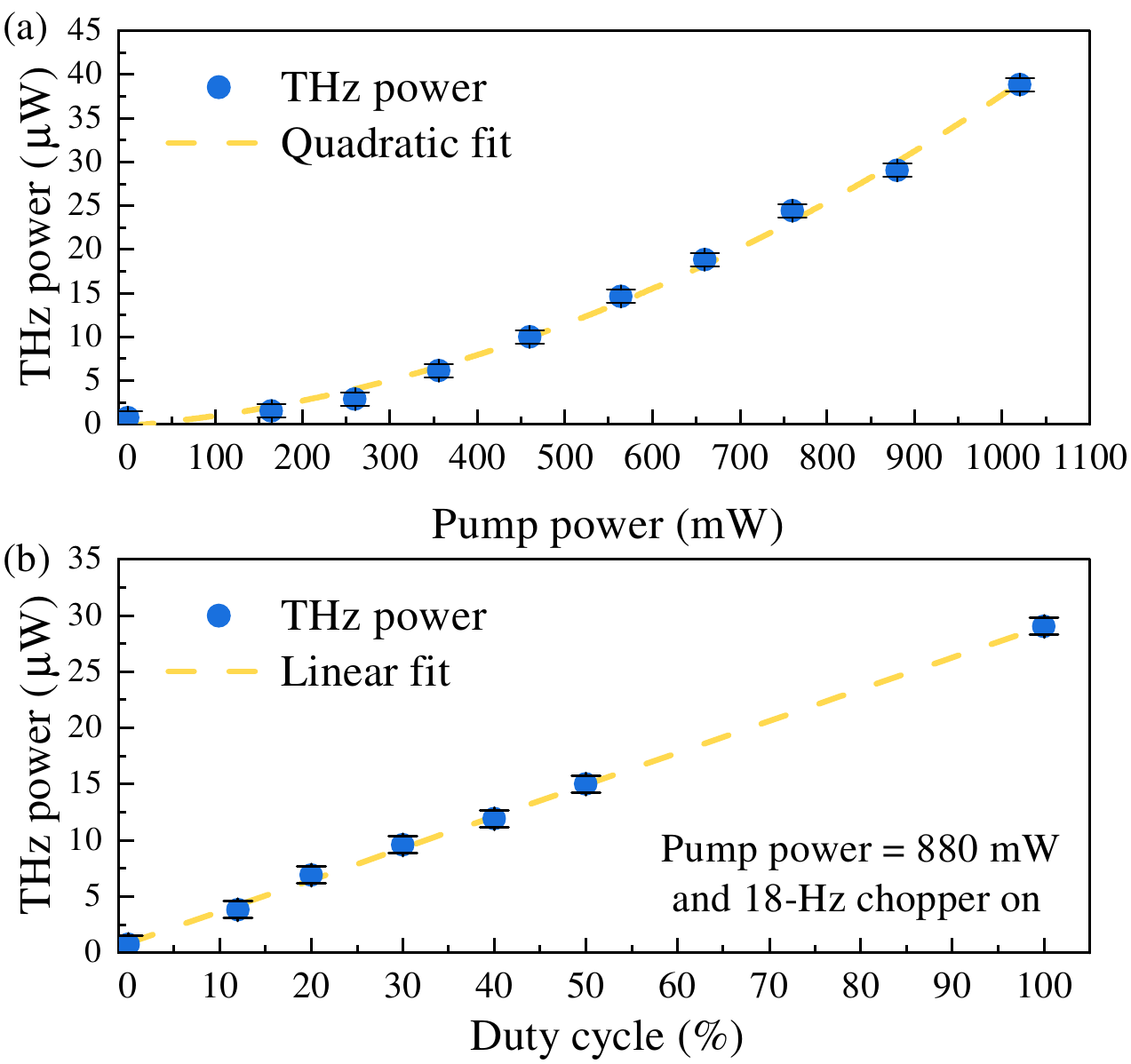}
\caption{(a) THz average power measured with the pyroelectric power meter as a function of the infrared pump power (solid points). The data fit with a quadratic curve (dashed line). (b) THz average power as a function of the second chopper's duty cycle (solid points). The data fit with a linear curve (dashed line). The infrared pump power is kept constant to 880~mW with the 18-Hz chopper on.}
\label{fig:power_and_duty}
\end{figure}
We measured the generated THz average power as a function of the pumping infrared power. Results are shown in Fig.\ref{fig:power_and_duty}(a).
The power out of the fiber amplifier sent to the DSTMS crystal is controlled with the HWP and PBS placed at the output of the amplified MO. 
The pump power is changed from a few mW to 1.02~W.
For all measurements, the chopper rotates at 18~Hz with 50\% duty cycle, halving the effective average power without affecting the pulse energy, which reached \SI{120}{\nano\joule} at maximum.
The maximum THz power generated is \SI{40}{\micro\watt}, corresponding to a peak electric field of \SI{1.63}{\kilo\volt\per\meter} at the focus. As expected by an OR process, the trend is purely quadratic as a function of the pump power.
The infrared pulse is pre-chirped of $\sim -600$ fs$^2$ to empirically maximize THz generation.
Efficiency saturation owing to thermal load is an important element that must be taken into account when using organic crystals. For instance, Ref.\cite{Mansourzadeh2021} shows a strong saturation of a BNA crystal for a duty cycle $>$20\% and an effective repetition rate of 2.7~MHz.
To investigate whether heat accumulation limits THz generation, we measured the average THz power as a function of the chopper duty cycle. The results are shown in Fig.\ref{fig:power_and_duty}(b). 
A second chopper, with variable blade coverage and high rotation speed (up to 1 kHz), was inserted into the beam path to modulate the duty cycle between 10\% and 50\%, effectively varying the duration of burst of infrared pulses. This way, the crystal experiences different ratios of generation time and cooling time for heat dissipation.
In numbers, for a cycle of 1~ms, at 10\% duty cycle, the resting time is 900~μs, while at 50\% duty cycle it is 500~μs.
The pump was kept constant at 880~mW with the slow 18-Hz chopper on, and the energy per pulse remained unchanged across all measurements. The the 18-Hz chopper does not affect the measurement, since is rotation has a time scale 100-times slower than the thermal dynamics in DSTMS.
At 50\% duty cycle, the THz generated is $\sim$15~µW.
The point of highest power is acquired by removing the second chopper (equivalent to a duty cycle of 100\%), obtaining $\sim \SI{30}{\micro\watt}$ of THz. For comparison, we generated THz using a conventional inorganic 500-µm-thick GaP crystal, AR-coated for the pump wavelength (EKSMA Optics).
In this case, the maximum average power measured after the OAPs with the pyroelectric power meter is \SI{2.3}{\micro\watt}, 17.4 times lower with respect to the DSTMS.
\begin{figure}[]
  \centering
  \begin{minipage}{\columnwidth}
    \centering
    \includegraphics[width=\linewidth]{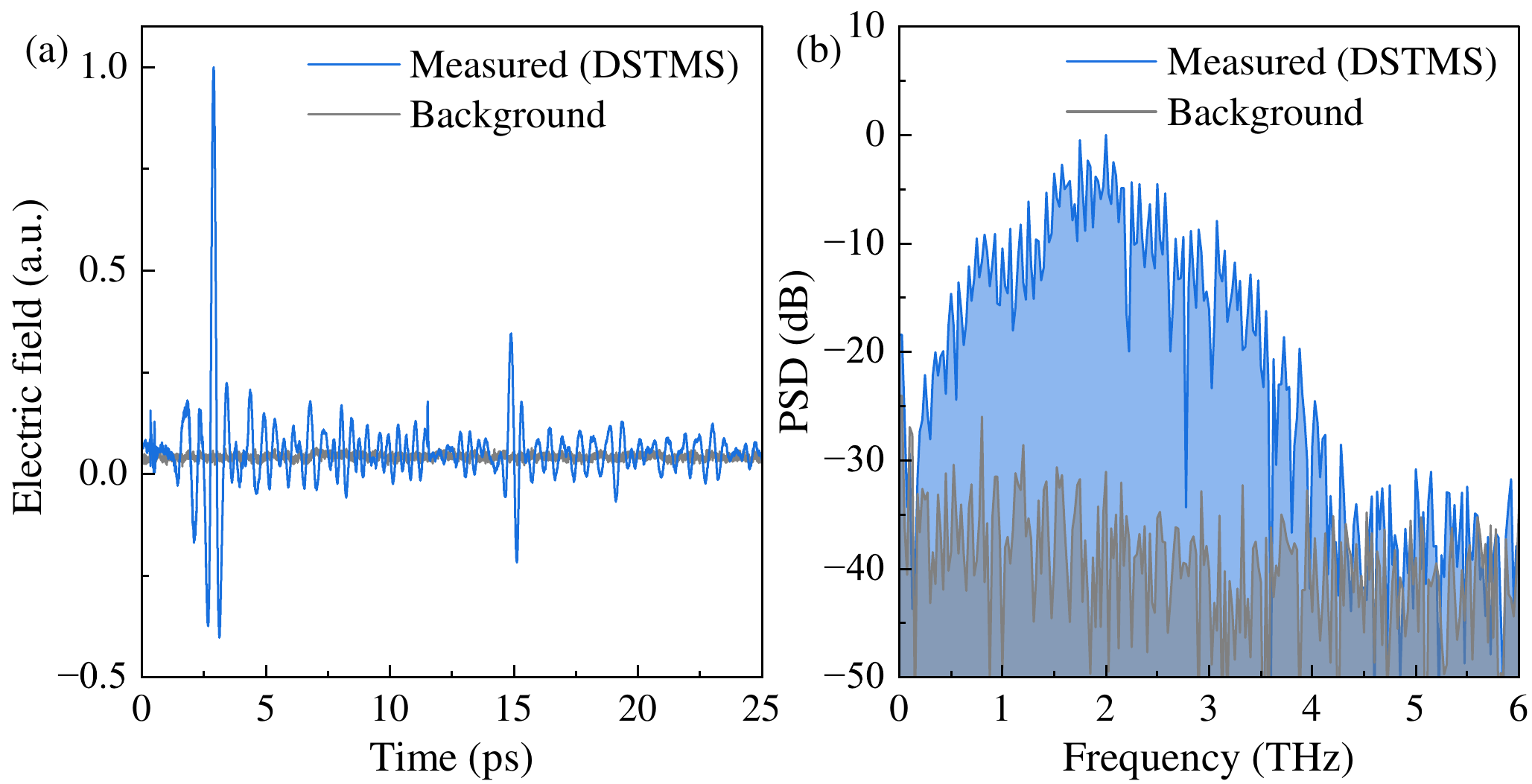}
  \end{minipage}

  \vspace{-3pt} 

  \begin{minipage}{\columnwidth}
    \centering
    \includegraphics[width=\linewidth]{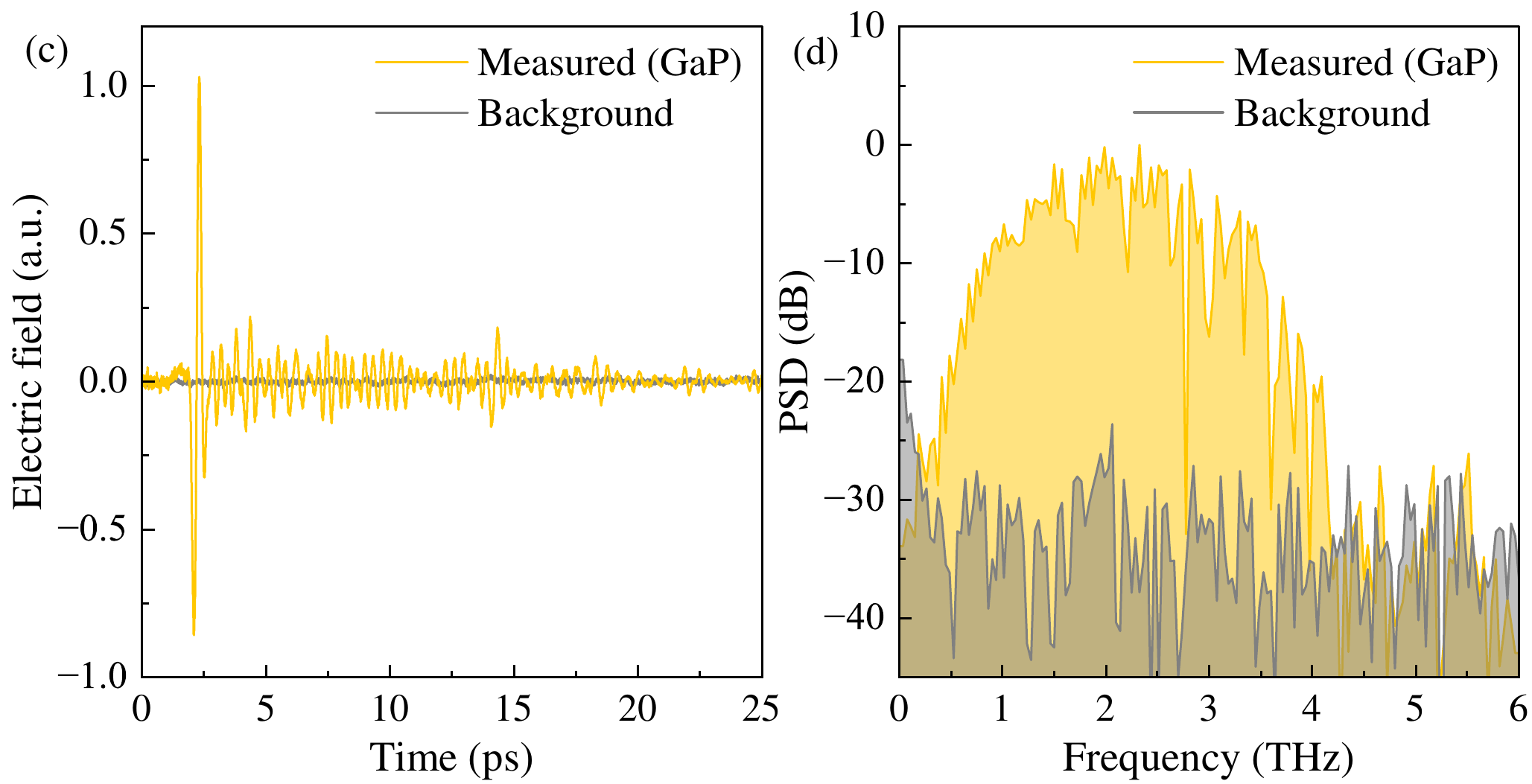}
  \end{minipage}

  \caption{Electro optical sampling of THz pulses and spectrum. Panels (a) and (c): Temporal traces of THz pulse generated with DSTMS and GaP, respectively Panels (b) and (d): Power spectral density of the THz pulses generated with DSTMS and GaP, respectively.}
  \label{fig:EOS_and_spectrum}
\end{figure}
As a second set of characterization measurements, we acquired the EOS trace of the THz pulses.
The EOS is performed by taking less then 10\% of power from the PBS at the output of the amplified MO.
This laser beam goes to a motorized delay line (MDL), then it is focused (with a lens of 200-mm focal length) and recombined with the THz pulses in a AR-coated 500-µW GaP crystal. The crystal length used is the shortest available in the lab, selected to optimize phase-matching and thereby the detection bandwidth \cite{Drs2019}. Nevertheless, we estimate that the THz generated with DSTMS is slightly attenuated above 4 THz due to this limitation.
When the laser and THz fields are properly overlapped and oriented in the crystal, the Pockels effect causes a change in the EOS crystal's refractive index, inducing a slight ellipticity of the laser beam polarization making the THz electric field detectable.
Both fields are linearly polarized at the interaction point.
A quarter-wave plate (QWP) and a Wollaston prism separate the two polarization components, which are sent to a balanced photodiode.
The output of the photodiode is proportional to the THz field synchronized with the infrared beam.
By scanning the delay between the THz and the probe beam, the full temporal waveform can be acquired with an oscilloscope.
To ensure synchronization between THz and probe beam, the beam used for OR travels a delay line (see Fig.\ref{fig:setup}).
Data are acquired using a lock-in amplifier (Stanford Research Systems SR530) with 1~s integration time, connected to an oscilloscope.
In this case, the chopper modulates the THz at \SI{10}{\kilo\hertz} while triggering the lock-in.
The same computer synchronizes the MDL and the data acquisition.
The MDL slowly drifts at a speed of 10 µm/s in a range of a few millimeters, corresponding to tens of ps of optical delay.
The EOS traces of the THz pulses generated by the DSTMS and the comparison GaP are shown in Fig.\ref{fig:EOS_and_spectrum}, respectively in panels (a) and (c).
Both traces show the few-cycle oscillation of the THz field, followed by long oscillations typical of the atmospheric water vapor absorption.
For both traces, a replica of the main pulse is visible at around 15~ps.
This is the etalon effect of the THz pulse inside the sampling GaP and is mainly due to the AR-coating for the infrared pump and more pronounced for the DSTMS measurement.
From the temporal traces it is possible to extract the power spectral density (PSD) via the Fourier transform. The spectrum of the THz pulses generated with DSTMS and GaP are shown in logarithmic scale in panels (b) and (d), respectively. 
Both spectra span from quasi-DC frequencies (below 500~GHz) to above 4~THz.
The spectrum of THz from GaP has a more regular shape, while THz from DSTMS shows a stronger modulation with $\sim$ 1.5~THz period and at 75~GHz, due to the aforementioned parasitic etalon effect in the sampling crystal.
The spectrum of THz from DSTMS extends below 250 GHz at low frequencies, and some power is visible at around 5~THz, after the phase-matching bounce to zero signal at 4.3~THz.
For both spectra, the SNR exceeds 30~dB.
The water absorption peaks are clearly visible in both spectra, with strong lines at 1.7~THz, 2.2~THz, 2.6 THz, 2.8~THz, 3.1~THz, and 3.6~THz.
The comparison between the two spectra shows no significant differences between the performances of DSTMS and GaP, despite a $\sim$20-fold difference in THz power, in good agreement with the efficiency ratio of a factor of $\sim$30 \cite{Mansourzadeh2023}.
This demonstrates  the strong potential of amplified Mamyshev oscillators for OR in organic crystals at MHz-repetition rate, opening promising perspectives for compact time-domain spectrometers based on this technology.
\begin{figure}[t!]
\centering
\includegraphics[width=\linewidth]{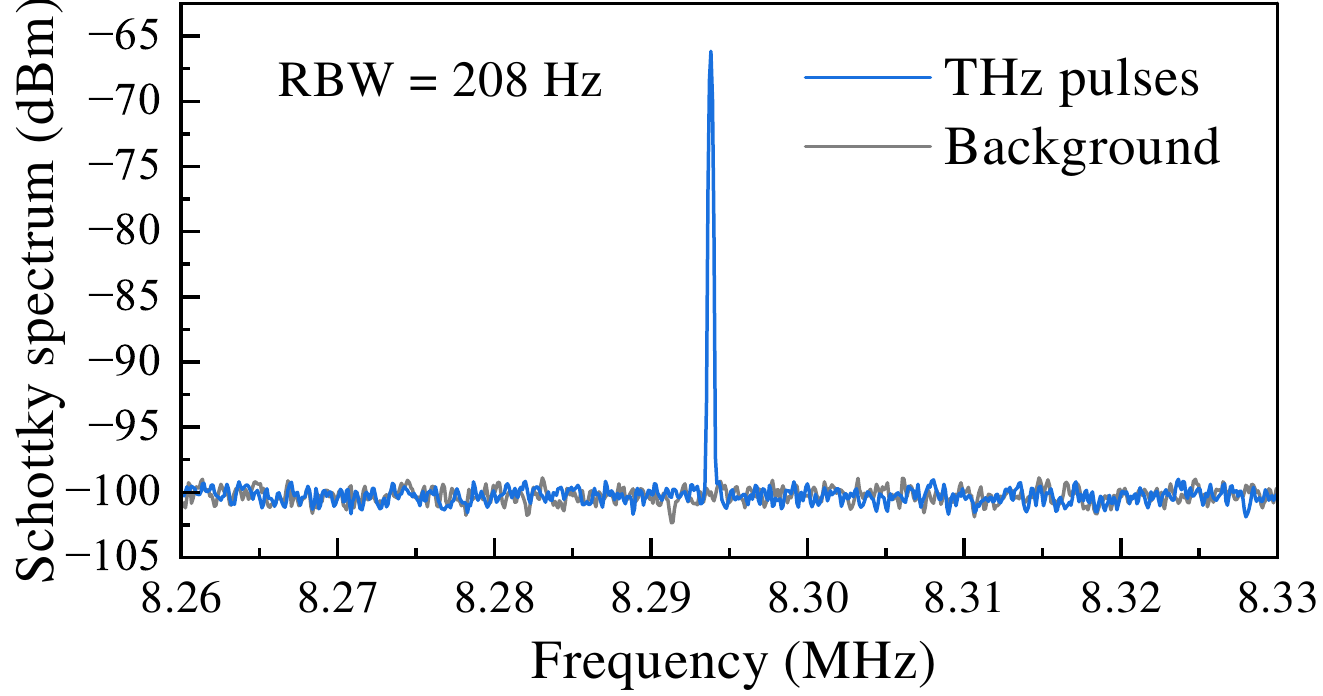}
\caption{Power spectrum of THz radiation from DSTMS acquired with Schottky diode. The resolution bandwidth (RBW) is 208 Hz.}
\label{fig:Schottky}
\end{figure}
As last measurement, we acquired the power spectrum of THz pulses from DSTMS around the repetition rate, using a quasi-optical Schottky diode (3DL-12C-LS2500-A1, ACST) and an electrical spectrum analyzer (Agilent E4445A).
The Schottky diode has an electrical bandwidth of 1~MHz, and optical bandwidth of a few THz.
The diode directly measures THz pulses and it is blind in the infrared (no chopper or lock-in are required).
Figure \ref{fig:Schottky} shows the repetition rate peak at 8.3~MHz measured with a resolution bandwidth of 208~Hz.
The SNR is high ($\sim$33~dB) despite we are measuring at frequencies beyond the detector electrical bandwidth (attenuation $\sim$20~dB). 
Apart from the sharp peak of the repetition rate at 8.3~MHz, no additional noise is observed above the detection limit, showing no significant noise degradation in the OR process.
This result is in agreement with the Mamyshev's low-noise performance reported in\cite{Pizzurro2025}.

In conclusion, we demonstrated THz pulse generation at 8.3~MHz via OR in a 190-µm DSTMS crystal, driven by a 31-fs, 1.1-W amplified Mamyshev oscillator. We achieved a maximum THz average power of 40 µW with a spectral bandwidth exceeding 4 THz, few-cycle temporal profiles, and no significant noise degradation. Compared to a 500-µm GaP crystal, DSTMS yielded almost 20-times higher output power with similar bandwidth, confirming the advantage of organic crystals for efficient, high-repetition-rate, broadband THz generation. This work opens promising prospects for Mamyshev-based compact, THz time-domain spectrometers operating at multi-MHz rates.

\subsection*{Acknowledgment}
This work was supported by: the European Union’s NextGenerationEU Programme with the I-PHOQS Infrastructure [IR0000016, ID D2B8D520, CUP B53C22001750006]; ATTILA - Advanced room-Temperature THz hyperspectral Imaging based on novel ultrafast fiber LAsers (20227849RL); and INFN CSN5, projects ETHIOPIA and ATHENAE.

\subsection*{Disclosures} The authors declare no conflicts of interest.

\subsection*{Data Availability Statement}
Data underlying the results presented in this paper are not publicly available at this time but may be obtained from the authors upon reasonable request.

\bibliography{bibliography}




\end{document}